\documentclass[aps,prd,reprint,preprintnumbers,nofootinbib,amsmath,amssymb,floatfix]{revtex4-2}

\usepackage{graphicx}
\usepackage{mathtools}
\usepackage[colorlinks=true,linkcolor=blue,citecolor=blue,urlcolor=blue]{hyperref}
\hypersetup{
  pdftitle={Yukawa-assisted charged-lepton dipoles in resonant electromagnetic leptogenesis},
  pdfauthor={Rin Takada}
}

\newcommand{\dd}{\mathrm d}
\newcommand{\mEM}{\tilde m^{\rm EM}}
\newcommand{\kmatch}{\kappa_{\rm match}}
\newcommand{\kN}{\kappa_N}
\newcommand{\kEW}{\kappa_{\rm EW}}
\newcommand{\order}{\mathcal O}

\newcommand{\MtildeBenchmark}{3.97\times10^{-2}}
\newcommand{\VacuumWidthBenchmark}{1.04\times10^{-10}}
\newcommand{\CNBHardBenchmark}{1.041\times10^{-12}}
\newcommand{\CNWHardBenchmark}{1.809\times10^{-12}}
\newcommand{\RhoBWBenchmark}{1.739}
\newcommand{\CNBMatchDisplay}{1.043\times10^{-12}}
\newcommand{\CNWMatchDisplay}{1.827\times10^{-12}}
\newcommand{\LogThree}{1.0986}
\newcommand{\SWOneTeV}{0.4901}
\newcommand{\CWOneTeV}{0.8717}
\newcommand{\SWWidth}{0.4803}
\newcommand{\CWWidth}{0.8771}
\newcommand{\MWWidth}{80.196}
\newcommand{\MZWidth}{91.431}
\newcommand{\RhoBlind}{1.7786}
\newcommand{\RhoNuBlind}{-1.8263}
\newcommand{\BlindDistancePercent}{2.24}
\newcommand{\XiGeneral}{0.8444}
\newcommand{\RhoStar}{-0.1903}
\newcommand{\XiReportedReal}{3.81\times10^{-5}}
\newcommand{\XiReportedPhase}{0.2983}
\newcommand{\FlavorE}{0.50}
\newcommand{\FlavorMu}{0.32}
\newcommand{\FlavorTau}{0.18}
\newcommand{\YNormBound}{1.62\times10^{-7}}
\newcommand{\TypeIBound}{1.59\times10^{-3}}
\newcommand{\QEDMuon}{0.926}
\newcommand{\QEDElectron}{0.890}
\newcommand{\QEDMuonAlt}{0.930}
\newcommand{\QEDElectronAlt}{0.897}
\newcommand{\QEDMuonShift}{0.35}
\newcommand{\QEDElectronShift}{0.79}
\newcommand{\RhoRoundedQuotient}{1.7378}
\newcommand{\MtildeReconstructed}{0.03965}
\newcommand{\RGEMatrixAA}{1.015078}
\newcommand{\RGEMatrixAB}{-0.0073333}
\newcommand{\RGEMatrixBA}{-0.0024444}
\newcommand{\RGEMatrixBB}{1.011220}
\newcommand{\RGEPositiveNB}{1.0431\times10^{-12}}
\newcommand{\RGEPositiveNW}{1.8272\times10^{-12}}
\newcommand{\RGEHalfPiNB}{1.0564\times10^{-12}}
\newcommand{\RGEHalfPiNW}{1.8297\times10^{-12}}
\newcommand{\RGEPiNB}{1.0696\times10^{-12}}
\newcommand{\RGEPiNW}{1.8323\times10^{-12}}
\newcommand{\LogSplitCorrection}{2.13\times10^{-8}}
\newcommand{\CrossingAmplitudeShift}{3.02\times10^{-9}}
\newcommand{\CrossingBRShift}{6.03\times10^{-9}}
\newcommand{\PeakAmplitudeShift}{2.45\times10^{-15}}
\newcommand{\DecayOnlyPeakSplitting}{1.29\times10^{-11}}
\newcommand{\DecayPlusScatteringPeakSplitting}{1.83\times10^{-11}}
\newcommand{\DecayOnlyPeakYield}{5.50\times10^{-5}}
\newcommand{\DecayPlusScatteringPeakYield}{8.44\times10^{-5}}
\newcommand{\DecayOnlyObservedSplitting}{1.68\times10^{-5}}
\newcommand{\DecayPlusScatteringObservedSplitting}{2.34\times10^{-5}}
\newcommand{\GeneralBR}{2.77\times10^{-28}}
\newcommand{\GeneralBRRatio}{1.85\times10^{-15}}
\newcommand{\GeneralEDM}{5.71\times10^{-37}}
\newcommand{\GeneralEDMRatio}{1.39\times10^{-7}}
\newcommand{\GeneralAMu}{1.31\times10^{-21}}
\newcommand{\GeneralAMuRatio}{9.06\times10^{-12}}
\newcommand{\FixedBR}{4.90\times10^{-29}}
\newcommand{\FixedEDM}{2.40\times10^{-37}}
\newcommand{\FixedAMu}{4.42\times10^{-22}}
\newcommand{\RealBR}{6.26\times10^{-33}}
\newcommand{\RealEDM}{2.71\times10^{-39}}
\newcommand{\RealAMu}{5.00\times10^{-24}}
\newcommand{\EndpointBR}{1.76\times10^{-25}}
\newcommand{\EndpointBRRatio}{1.18\times10^{-12}}
\newcommand{\EndpointEDM}{1.44\times10^{-35}}
\newcommand{\EndpointEDMRatio}{3.51\times10^{-6}}
\newcommand{\EndpointAMu}{3.31\times10^{-20}}
\newcommand{\EndpointAMuRatio}{2.29\times10^{-10}}

\begin{document}

\preprint{RESCEU-30/25}

\title{Yukawa-assisted charged-lepton dipoles in resonant electromagnetic leptogenesis}

\author{Rin Takada}
\email{takada-rin@resceu.s.u-tokyo.ac.jp}
\affiliation{Research Center for the Early Universe (RESCEU), Graduate School of Science,
The University of Tokyo, 7-3-1 Hongo, Bunkyo, Tokyo 113-0033, Japan}
\date{\today}

\begin{abstract}
We study the charged-lepton dipole contribution that is linear in the
neutrino-dipole coefficients $C_{NB,NW}$ and in the renormalizable neutrino
Yukawa matrix $Y_\nu$.  We consider a resonant EMLG benchmark with dipole-sector
coefficients renormalized at $\kN=M_1=1\,\mathrm{TeV}$ and evolved through the
coupled one-loop RGE to $\kmatch=3\,\mathrm{TeV}$, a factorized flavor structure,
and $\mEM_1=\MtildeBenchmark\,\mathrm{eV}$.  Since these inputs do not determine
$Y_\nu$, we impose
the conditional width bound $\Gamma_Y^{(0)}\leqslant\epsilon\Gamma_{\rm EM}^{(0)}$
and a vanishing direct ultraviolet charged-lepton dipole, derive the one-loop
$\nu$SMEFT leading logarithm, and apply the tree-level electroweak projection
and one-loop QED evolution in LEFT\@.  For $\epsilon=10^{-2}$ and two coherently
aligned heavy states, the width-only envelopes are
$\mathrm{BR}(\mu\to e\gamma)\leqslant\GeneralBR$,
$|d_e|\leqslant\GeneralEDM\,e\,\mathrm{cm}$, and
$|\Delta a_\mu|\leqslant\GeneralAMu$.  The factorized flavor structure and
$C_{NW}/C_{NB}=\RhoBWBenchmark$ give smaller conditional bounds.  The benchmark
phase $\phi_{BW}=0$, for which $\rho_{BW}=+\RhoBWBenchmark$, lies near the
charged-lepton photon-dipole blind direction.
The vacuum local coefficient contains no resonant pole denominator.  The two
thermal pole prescriptions enter the low-energy comparison only through the
mass splittings selected by the transport calculation, producing a relative
change of order $\Delta M/M_1$.
\end{abstract}

\maketitle

\section{Introduction}
\label{sec:introduction}

Electroweak sphalerons convert a lepton asymmetry into baryon number when the
Sakharov conditions are realized before sphaleron freeze-out
\cite{Sakharov1967,KlinkhamerManton1984,FukugitaYanagida1986,DavidsonNardiNir2008}.
Electromagnetic leptogenesis (EMLG) replaces the usual heavy-neutrino Yukawa
decay vertex with gauge-invariant transition-dipole operators
\cite{BellKayserLaw2008}.  TeV-scale realizations require the renormalizable
neutrino Yukawa interaction to be suppressed relative to the dipole channel
\cite{ChoudhuryEtAl2012}.

The companion analysis \cite{TakadaHelicity} defines a two-state resonant EMLG
benchmark with helicity-resolved transport, a finite-rate sphaleron equation,
and two pole prescriptions.  It specifies the neutrino-dipole coefficients,
their factorized flavor structure, the physical vacuum width, and the scales at
which these inputs are renormalized.  The present calculation addresses the
vacuum EFT below the heavy-neutrino thresholds and isolates the
charged-lepton dipole term proportional to one neutrino-dipole coefficient
and one power of $Y_\nu$.

The calculation retains the one-loop leading logarithm and applies the QED
evolution below the electroweak scale as an RG improvement.  Finite one-loop
matching at the heavy-neutrino thresholds is not included.
Charged-lepton dipoles containing only $C_{NX}$ and no $Y_\nu$ require
additional loops and lie outside this linear analysis.

Section~\ref{sec:eft} fixes the EFT conventions, scales, running, and physical
width.  Section~\ref{sec:benchmark} extracts the benchmark inputs and the
mass-splitting information relevant to the low-energy comparison.
Sections~\ref{sec:mixing}
and~\ref{sec:projection} derive the linear source and its electroweak
projection.  Section~\ref{sec:observables} gives the LEFT evolution and
observable normalizations.  The conditional envelopes are presented in
Sec.~\ref{sec:results}, followed by the conclusions in
Sec.~\ref{sec:discussion}.

\section{EFT setup and matching scales}
\label{sec:eft}

\subsection{Neutrino dipoles and broken-phase couplings}

Above the right-handed-neutrino thresholds, the relevant $\nu$SMEFT operators
\cite{BellKayserLaw2008,Datta2021,ArduMarcano2024} are
\begin{align}
\mathcal O_{NB,\alpha i}
 &\coloneqq(\bar L_\alpha\sigma^{\mu\nu}P_RN_i)\tilde H B_{\mu\nu},
\label{eq:ONB}\\
\mathcal O_{NW,\alpha i}
 &\coloneqq(\bar L_\alpha\sigma^{\mu\nu}\tau^AP_RN_i)
    \tilde H W^A_{\mu\nu},
\label{eq:ONW}
\end{align}
where $\tau^A$ are the Pauli matrices, not $\sigma^A/2$, and
\begin{equation}
\sigma^{\mu\nu}=\frac{\mathrm{i}}{2}[\gamma^\mu,\gamma^\nu].
\end{equation}
The effective Lagrangian is
\begin{equation}
-\mathcal L_{\nu{\rm SMEFT}}
\supset
C_{NB,\alpha i}\mathcal O_{NB,\alpha i}
+C_{NW,\alpha i}\mathcal O_{NW,\alpha i}
+\mathrm{h.c.},
\label{eq:nusmeft-lagrangian}
\end{equation}
with $[C_{NX}]=-2$.  If the weak operator is instead defined with
$T^A=\sigma^A/2$, the corresponding coefficients satisfy
$C_{NW}=C_{NW}^{(T)}/2$.

For $H=(0,(v_0+h)/\sqrt2)^{\intercal}$ and
$\tilde H=((v_0+h)/\sqrt2,0)^{\intercal}$, the zero-temperature broken-phase
couplings are
\begin{align}
\mu^B_{\alpha i}&=\frac{v_0}{\sqrt2}C_{NB,\alpha i},&
\mu^3_{\alpha i}&=\frac{v_0}{\sqrt2}C_{NW,\alpha i},
\notag\\
\mu^W_{\alpha i}&=v_0C_{NW,\alpha i},&
\mu^\gamma_{\alpha i}&=c_0\mu^B_{\alpha i}+s_0\mu^3_{\alpha i},
\notag\\
\mu^Z_{\alpha i}&=-s_0\mu^B_{\alpha i}+c_0\mu^3_{\alpha i}.
\label{eq:broken-couplings}
\end{align}
Thus $\mu^W=\sqrt2\,\mu^3$ in the Pauli-matrix convention, which produces
the factor of two in the charged-current contribution to the width.  Here
$(s_0,c_0)$ denotes the zero-temperature weak-angle convention used to
normalize the companion benchmark width.  It is distinguished below from the
strict-leading-logarithmic angle evaluated at $\kN$.

The vacuum EFT sequence is
\begin{equation}
\begin{gathered}
\nu{\rm SMEFT}\quad(\kmatch>\kappa>M_i)
\longrightarrow {\rm SMEFT},\\
{\rm SMEFT}\quad(M_i>\kappa>\kEW)
\longrightarrow {\rm LEFT}\quad(\kappa<\kEW),
\end{gathered}
\label{eq:eft-chain}
\end{equation}
where $\kEW=150\,\mathrm{GeV}$ is an electroweak matching reference and is
not the thermal crossover temperature used in the transport calculation.

\subsection{Coupled evolution between \texorpdfstring{$\kN$ and $\kmatch$}{kappaN and kappamatch}}

In the convention of Eqs.~\eqref{eq:ONB} and~\eqref{eq:ONW}, the gauge terms
and top-Yukawa trace give \cite{Datta2021,ArduMarcano2024}
\begin{align}
16\pi^2\kappa\frac{\dd C_{NB}}{\dd\kappa}
={}&\left(\frac{91}{12}g'^2-\frac94g^2+3y_t^2\right)C_{NB}
\notag\\[-2pt]
&-\frac92gg'C_{NW},
\label{eq:RGE-NB}\\
16\pi^2\kappa\frac{\dd C_{NW}}{\dd\kappa}
={}&\left(-\frac34g'^2-\frac{11}{12}g^2+3y_t^2\right)C_{NW}
\notag\\[-2pt]
&-\frac32gg'C_{NB}.
\label{eq:RGE-NW}
\end{align}
Flavor indices are suppressed.  Charged-lepton, down-quark, and
renormalizable-neutrino-Yukawa terms are omitted, as in the companion
benchmark.

The benchmark inputs at the heavy scale are
\begin{align}
\kN=M_1&=1\,\mathrm{TeV},\qquad \kmatch=3\,\mathrm{TeV},
\label{eq:scales}\\
|C_{NB,e1}(\kN)|&=\CNBHardBenchmark\,\mathrm{GeV}^{-2},
\notag\\[-2pt]
|C_{NW,e1}(\kN)|&=\CNWHardBenchmark\,\mathrm{GeV}^{-2}.
\label{eq:C-kN}
\end{align}
The same unrounded benchmark input defines the ratio
\begin{equation}
\frac{C_{NW}(\kN)}{C_{NB}(\kN)}=+\RhoBWBenchmark
\label{eq:rho-modulus}
\end{equation}
as a real-positive EFT input at $\kN$.
All calculations use one unrounded value of $C_{NB,e1}$ and one unrounded
ratio, with $C_{NW,e1}$ generated from their product.  The quotient
$\RhoRoundedQuotient$ reconstructed from the two independently rounded
numbers in Eq.~\eqref{eq:C-kN} need not reproduce the displayed ratio exactly.
We write
\begin{equation}
\rho_{BW}\coloneqq\frac{C_{NW}}{C_{NB}}
=|\rho_{BW}|\mathrm{e}^{\mathrm{i}\phi_{BW}},
\qquad |\rho_{BW}|=\RhoBWBenchmark.
\label{eq:rho-phase}
\end{equation}
The real-positive input in Eq.~\eqref{eq:rho-modulus} fixes the benchmark
phase to $\phi_{BW}=0$, for which $\rho_{BW}=+\RhoBWBenchmark$; relaxed phases at fixed
modulus are used below only as sensitivity slices.

Because Eqs.~\eqref{eq:RGE-NB} and~\eqref{eq:RGE-NW} mix the two complex
coefficients, the endpoint magnitudes depend on their relative phase.  The
one-loop SM running and numerical inputs are specified in
Appendix~\ref{app:inputs}.  Writing
$\mathbf C_N=(C_{NB},C_{NW})^{\intercal}$, the evolution is
\begin{align}
\mathbf C_N(\kmatch)&=U_{3/1}\,\mathbf C_N(\kN),
\notag\\[-2pt]
U_{3/1}&=
\begin{pmatrix}
\RGEMatrixAA&\RGEMatrixAB\\
\RGEMatrixBA&\RGEMatrixBB
\end{pmatrix}.
\label{eq:RGE-transfer}
\end{align}
For the benchmark phase $\phi_{BW}=0$, the evolved magnitudes are
$|C_{NB}|=\RGEPositiveNB\,\mathrm{GeV}^{-2}$ and
$|C_{NW}|=\RGEPositiveNW\,\mathrm{GeV}^{-2}$ at $\kmatch$.
The companion manuscript separately quotes
\begin{align}
|C_{NB,e1}(\kmatch)|&=\CNBMatchDisplay\,\mathrm{GeV}^{-2},
\notag\\[-2pt]
|C_{NW,e1}(\kmatch)|&=\CNWMatchDisplay\,\mathrm{GeV}^{-2}.
\label{eq:C-kmatch}
\end{align}
Equation~\eqref{eq:C-kmatch} is the three-significant-digit form of the
coupled evolution of the common unrounded benchmark.  Other phases would
give different endpoint magnitudes; examples are
listed in Appendix~\ref{app:inputs}.

The coupled evolution is used only to relate renormalized benchmark inputs at
the two scales.  In the linear one-loop leading-logarithmic source below,
$C_{NX}$ is held fixed.  Substituting its one-loop running into the same
one-loop integral produces a term of order
$(16\pi^2)^{-2}\ln^2(\kmatch/M_i)$, which is a two-loop leading logarithm.

\subsection{Physical widths and the conditional Yukawa bound}

For $V=Z,W$, define the vacuum phase-space factor
\begin{equation}
R_{V,i}^{\rm vac}
\coloneqq\Theta(1-r_{V,i})(1-r_{V,i})^2
 \left(1+\frac{r_{V,i}}{2}\right),
\qquad r_{V,i}\coloneqq\frac{m_V^2}{M_i^2},
\label{eq:phase-space}
\end{equation}
with $R_{\gamma,i}^{\rm vac}\coloneqq1$.  Neglecting the light-lepton mass, the
charge-conjugate-summed Majorana width is
\begin{equation}
\Gamma^{(0)}_{V,\alpha i}
=\frac{M_i^3}{2\pi}R_{V,i}^{\rm vac}|\mu^V_{\alpha i}|^2.
\label{eq:majorana-width}
\end{equation}
Appendix~\ref{app:widths} gives the phase-space and Majorana factors.

The physical-width parameter is
\begin{align}
\mEM_i
&\coloneqq v_0^2M_i\sum_{\alpha,V}R_{V,i}^{\rm vac}|\mu^V_{\alpha i}|^2,
\notag\\[-2pt]
\Gamma_{{\rm EM},i}^{(0)}\coloneqq\Gamma_i^{\rm vac}
&=\frac{M_i^2}{2\pi v_0^2}\mEM_i,
\label{eq:mtilde-definition}
\end{align}
where $v_0=246\,\mathrm{GeV}$, all couplings are zero-temperature
broken-phase quantities, and $V=\gamma,Z,W$.

The renormalizable interaction is
\begin{equation}
-\mathcal L_Y\supset
(Y_\nu)_{\alpha i}\bar L_\alpha\tilde H N_i+\mathrm{h.c.},
\end{equation}
with the charge-conjugate-summed width\footnote{The expression below uses
massless final states.  Including the broken-phase $W$, $Z$, and $h$
phase-space factors changes the summed Yukawa width by less than $1\%$ for
$M_i=1\,\mathrm{TeV}$.}
\begin{equation}
\Gamma_{Y,i}^{(0)}
=\frac{M_i}{8\pi}(Y_\nu^\dagger Y_\nu)_{ii}.
\label{eq:yukawa-width}
\end{equation}
To define a nonzero Yukawa extension while keeping the dipole channel
dominant in the vacuum two-body width, we impose
\begin{equation}
\Gamma_{Y,i}^{(0)}\leqslant\epsilon\Gamma_{{\rm EM},i}^{(0)},
\label{eq:width-inequality}
\end{equation}
which gives
\begin{equation}
(Y_\nu^\dagger Y_\nu)_{ii}
\leqslant 4\epsilon\frac{M_i}{v_0^2}\mEM_i.
\label{eq:Y-norm-bound}
\end{equation}
This is an additional condition on vacuum widths.  It is not part of the
companion benchmark and does not bound all thermal scattering or washout
rates.

\section{Benchmark inherited from resonant EMLG}
\label{sec:benchmark}

\subsection{Factorized dipole texture and phase information}

The companion benchmark uses
\begin{equation}
C_{NX,\alpha i}=c_X y_{\Sigma,\alpha}y_{H,i}^{\ast},
\qquad X=B,W,
\label{eq:factorized}
\end{equation}
with
\begin{align}
y_H&=(1.4,\,1.4\mathrm{e}^{-0.7\mathrm{i}})^{\intercal},
\notag\\
y_\Sigma&=(1.4,\,1.12\mathrm{e}^{-0.3\mathrm{i}},
\,0.84\mathrm{e}^{-1.1\mathrm{i}})^{\intercal}.
\label{eq:factorized-vectors}
\end{align}
The phases are benchmark choices.  The vector $y_H$ is a dipole-texture
parameter and is not the renormalizable matrix $Y_\nu$.  The dipole flavor
fractions are
\begin{equation}
f_\alpha
\coloneqq\frac{|y_{\Sigma,\alpha}|^2}{\sum_\beta|y_{\Sigma,\beta}|^2},
\qquad
(f_e,f_\mu,f_\tau)=(\FlavorE,\FlavorMu,\FlavorTau).
\label{eq:flavor-fractions}
\end{equation}
Since $|y_{H,1}|=|y_{H,2}|$ and
$M_2/M_1=1+\order(\Delta M/M_1)$, the two states have equal widths at the
accuracy used for the low-energy logarithm.

The physical-width benchmark is
\begin{equation}
\mEM_1=\MtildeBenchmark\,\mathrm{eV},
\qquad
\Gamma_1^{\rm vac}=\VacuumWidthBenchmark\,\mathrm{GeV}.
\label{eq:benchmark-width}
\end{equation}
The common unrounded Wilson coefficients, the factorized flavor fractions,
and the zero-temperature width convention give
$\mEM_1=\MtildeReconstructed\,\mathrm{eV}$, which rounds to
Eq.~\eqref{eq:benchmark-width}.

The linear Yukawa-assisted coefficient requires $Y_\nu$, which is not
specified by the companion benchmark.  The numerical envelopes below therefore
use the conditional norm bound in Eq.~\eqref{eq:Y-norm-bound}.

\subsection{Mass splitting and pole prescriptions}

The decay-only and decay-plus-scattering pole prescriptions in
Ref.~\cite{TakadaHelicity} select different positive near-resonant maxima and
different positive observed-yield crossings.  Table~\ref{tab:poles} records
the corresponding mass splittings.

\begin{table}[tb]
\caption{Mass splittings selected by the companion transport calculation at
$\mEM_1=\MtildeBenchmark\,\mathrm{eV}$.  The last column lists the values of
$\Delta M$ at which $Y_B^{\rm FO}$ crosses the observed value
$8.7\times10^{-11}$.}
\label{tab:poles}
\squeezetable
\begin{ruledtabular}
\begin{tabular}{lccc}
Prescription & $\Delta M_{\rm peak}$ & Peak $Y_B^{\rm FO}$ & $\Delta M_{\rm obs}$ \\
& $[\mathrm{GeV}]$ && $[\mathrm{GeV}]$ \\
\hline
Decay only
& $\DecayOnlyPeakSplitting$ & $+\DecayOnlyPeakYield$ & $\DecayOnlyObservedSplitting$ \\
\begin{tabular}{@{}l@{}}Decay plus\\scattering\end{tabular}
& $\DecayPlusScatteringPeakSplitting$ & $+\DecayPlusScatteringPeakYield$
& $\DecayPlusScatteringObservedSplitting$
\end{tabular}
\end{ruledtabular}
\end{table}

The direct mass dependence of the pole-free local leading logarithm is
\begin{equation}
L_i\coloneqq\ln\frac{\kmatch}{M_i},
\qquad
L_2-L_1=-\frac{\Delta M}{M_1}
+\order\!\left(\frac{\Delta M^2}{M_1^2}\right).
\label{eq:log-splitting}
\end{equation}
At the positive observed-yield crossing with the larger mass splitting,
$|(L_2-L_1)/L_1|=\LogSplitCorrection$.  The phase-space factors have the same
analytic $\order(\Delta M/M_1)$ dependence.  The pole prescriptions modify the
near-on-shell thermal self-energy treatment and the locations selected in
Table~\ref{tab:poles}; they do not modify the vacuum anomalous dimension or
insert a resonant denominator into a local coefficient at $q^2=0$.
For constructively aligned heavy-state contributions, the relative changes in
the local amplitude and its squared magnitude between the two crossings are
only $\CrossingAmplitudeShift$ and $\CrossingBRShift$, respectively.  The
relative amplitude change between the two peak locations is
$\PeakAmplitudeShift$.  These effects are of order
$|(\Delta M^{\rm D+S}-\Delta M^{\rm D})/M_1|$ and lie beyond the retained
leading-logarithmic accuracy.

\section{Yukawa-assisted mixing}
\label{sec:mixing}

\subsection{One-loop source and heavy thresholds}

The charged-lepton SMEFT dipoles are \cite{Grzadkowski2010}
\begin{align}
\mathcal O_{eB,\alpha\beta}
&\coloneqq(\bar L_\alpha\sigma^{\mu\nu}E_\beta)H B_{\mu\nu},
\label{eq:OeB}\\
\mathcal O_{eW,\alpha\beta}
&\coloneqq(\bar L_\alpha\sigma^{\mu\nu}\tau^AE_\beta)H W^A_{\mu\nu}.
\label{eq:OeW}
\end{align}
The Yukawa-dependent one-loop terms in their dimensionful coefficients are
\cite{ArduMarcano2024}
\begin{align}
16\pi^2\frac{\dd(C_{eB})_{\alpha\beta}}{\dd\ln\kappa}
&=-(C_{NB})_{\alpha i}[Y_\nu^\dagger Y_e]_{i\beta}+\cdots,
\label{eq:mixing-B}\\
16\pi^2\frac{\dd(C_{eW})_{\alpha\beta}}{\dd\ln\kappa}
&=-(C_{NW})_{\alpha i}[Y_\nu^\dagger Y_e]_{i\beta}+\cdots.
\label{eq:mixing-W}
\end{align}
In the charged-lepton mass basis,
\begin{equation}
[Y_\nu^\dagger Y_e]_{i\beta}
=(Y_\nu)_{\beta i}^{\ast}y_\beta,
\qquad y_\beta\coloneqq\frac{\sqrt2m_\beta}{v_0}.
\label{eq:index-contraction}
\end{equation}
The complex conjugation follows directly from the Hermitian conjugate in
$Y_\nu^\dagger$ and fixes the phase entering a diagonal CP-odd coefficient.

To isolate the linear portal, we impose the additional UV boundary condition
\begin{equation}
C_{eB}^{\rm direct}(\kmatch)
=C_{eW}^{\rm direct}(\kmatch)=0.
\label{eq:direct-boundary}
\end{equation}
This condition is not implied by $\mEM_i$ or by the factorized neutrino-dipole
texture.

The source associated with $N_i$ terminates at its threshold.  Below both
thresholds,
\begin{align}
(C_{eX}^{\rm lin,LL})_{\alpha\beta}
={}&\frac{1}{16\pi^2}\sum_i
\int_{M_i}^{\kmatch}\frac{\dd\kappa}{\kappa}
(C_{NX})_{\alpha i}(\kappa)
\notag\\[-2pt]
&\hspace{22mm}\times[Y_\nu^\dagger Y_e]_{i\beta}(\kappa),
\label{eq:integrated-source}\\
={}&\frac{y_\beta}{16\pi^2}\sum_i L_i
(C_{NX})_{\alpha i}(Y_\nu)_{\beta i}^{\ast}
\notag\\[-2pt]
&+\text{higher logarithms},
\label{eq:strict-LL}
\end{align}
where $X=B,W$.  The positive logarithm follows by integrating the negative
anomalous dimension from $\kmatch$ down to $M_i$.

At leading-logarithmic accuracy, the quasi-degenerate pair can be integrated
out at a common threshold.  A finite fixed-order calculation should organize the
pair jointly in heavy-flavor space or decouple it sequentially with a
consistent basis prescription.  Explicit one-loop matching calculations in
the type-I seesaw illustrate the required finite-threshold bookkeeping
\cite{ZhangZhouRadiative2021,ZhangZhouComplete2021}, although the mixed
dipole--Yukawa threshold term in Eq.~\eqref{eq:strict-LL} has not been computed
here.

At the benchmark, $L_1=\ln 3=\LogThree$, which is not parametrically large.
Finite one-loop threshold terms can therefore be comparable to the logarithmic
term in the linear amplitude.  Near a tree-level electroweak projection zero,
the leading logarithm alone does not determine the complete one-loop linear
result.

\section{Electroweak projection and blind directions}
\label{sec:projection}

\subsection{Photon projection}

Below the electroweak scale, define
\begin{align}
\mathcal O_{e\gamma,\alpha\beta}
&\coloneqq(\bar\ell_\alpha\sigma^{\mu\nu}P_R\ell_\beta)F_{\mu\nu},
\notag\\
-\mathcal L_{\rm LEFT}&\supset
(C_{e\gamma})_{\alpha\beta}\mathcal O_{e\gamma,\alpha\beta}
+\mathrm{h.c.}.
\label{eq:LEFT-operator}
\end{align}
The tree-level electroweak projection is
\begin{equation}
(C_{e\gamma})_{\alpha\beta}(\kEW)
=\frac{v_0}{\sqrt2}
\left[c_N(C_{eB})_{\alpha\beta}
-s_N(C_{eW})_{\alpha\beta}\right].
\label{eq:EW-projection}
\end{equation}
The minus sign follows from the lower Higgs component in
$\mathcal O_{eW}$, whereas the neutrino photon dipole in
Eq.~\eqref{eq:broken-couplings} originates from the upper component of
$\tilde H$.  Here $(s_N,c_N)\coloneqq(s_W(\kN),c_W(\kN))$ is used in
the strict one-loop leading logarithm.  Replacing it by the weak angle at
$\kEW$ changes an already one-loop coefficient at relative one-loop order
and therefore contributes beyond this accuracy.

Define
\begin{equation}
\mu^{e\gamma}_{\alpha i}
\coloneqq c_N\mu^B_{\alpha i}-s_N\mu^3_{\alpha i}.
\label{eq:mu-egamma}
\end{equation}
The linear leading-logarithmic coefficient is
\begin{align}
(C_{e\gamma}^{\rm lin,LL})_{\alpha\beta}(\kEW)
&=\frac{y_\beta}{16\pi^2}P_{\alpha\beta},
\label{eq:Cegamma-P}\\
P_{\alpha\beta}
&\coloneqq\sum_i L_i(Y_\nu)_{\beta i}^{\ast}\mu^{e\gamma}_{\alpha i}.
\label{eq:P-definition}
\end{align}
No factor of $[(M_2^2-M_1^2)+\mathrm{i}M\Gamma]^{-1}$ occurs in
Eq.~\eqref{eq:P-definition}.  The resonant denominator describes near-on-shell
propagation in the thermal source, while Eq.~\eqref{eq:Cegamma-P} is an
analytic local coefficient at vanishing external momentum.

\subsection{\texorpdfstring{$B$--$W$}{B--W} directions}

For proportional $B$ and $W$ flavor vectors with a common complex ratio
$\rho=C_{NW}/C_{NB}$,
\begin{align}
\mu^\gamma&\propto c_0+s_0\rho,&
\mu^Z&\propto -s_0+c_0\rho,
\notag\\
\mu^W&\propto\sqrt2\rho,&
\mu^{e\gamma}&\propto c_N-s_N\rho.
\label{eq:directions}
\end{align}
The photon blind directions are
\begin{align}
\rho_{e\gamma}^{\rm blind}&=+\frac{c_N}{s_N},
\label{eq:charged-blind}\\
\rho_{\nu\gamma}^{\rm blind}&=-\frac{c_0}{s_0}.
\label{eq:neutrino-blind}
\end{align}
The first cancels the charged-lepton photon projection; the second cancels
$N\to\nu\gamma$.  In the two scale conventions used here, their numerical
locations are $\RhoBlind$ and $\RhoNuBlind$, respectively.  A UV relation
$\rho=g/g'$ would coincide with the first
only if the coefficient ratio and weak angle were evaluated at the same
scale.  Equation~\eqref{eq:rho-modulus} instead specifies an independent EFT
input; it is neither such a UV relation nor an RGE prediction.

The common SM inputs are evolved from $\kEW$ to $\kN$ with the one-loop
equations in Appendix~\ref{app:inputs}.  They give
\begin{equation}
s_N=\SWOneTeV,
\qquad c_N=\CWOneTeV,
\qquad \frac{c_N}{s_N}=\RhoBlind.
\label{eq:weak-numbers}
\end{equation}
The benchmark $\rho_{BW}=+\RhoBWBenchmark$ lies \BlindDistancePercent\% below the
charged-lepton blind direction at this scale.  Its small projected value is
therefore sensitive to higher-order scale evolution and finite threshold
matching.

\subsection{Width-normalized maximum}

For one heavy state, define
\begin{equation}
z_{\alpha i}\coloneqq
\begin{pmatrix}\mu^B_{\alpha i}\\[2pt]\mu^3_{\alpha i}\end{pmatrix},
\qquad
w_N\coloneqq\begin{pmatrix}c_N\\-s_N\end{pmatrix}.
\end{equation}
The contribution of the same flavor vector to the electromagnetic width is
$z_{\alpha i}^\dagger D_i^{(0)}z_{\alpha i}$, with
\begin{equation}
D_i^{(0)}\coloneqq
\begin{pmatrix}
 c_0^2+R_{Z,i}^{\rm vac}s_0^2
 &(1-R_{Z,i}^{\rm vac})c_0s_0\\
 (1-R_{Z,i}^{\rm vac})c_0s_0
 &s_0^2+R_{Z,i}^{\rm vac}c_0^2+2R_{W,i}^{\rm vac}
\end{pmatrix}_{M_i}.
\label{eq:D-matrix}
\end{equation}
This is the companion zero-temperature width convention, for which
$s_0=\SWWidth$, $c_0=\CWWidth$,
$m_W=\MWWidth\,\mathrm{GeV}$, and $m_Z=\MZWidth\,\mathrm{GeV}$.
For open channels, $D_i^{(0)}$ is positive definite.  The generalized
Cauchy--Schwarz inequality gives
\begin{align}
\sum_\alpha|\mu^{e\gamma}_{\alpha i}|^2
&\leqslant
\Xi_{e\gamma}^{\max}(M_i)
\frac{\mEM_i}{v_0^2M_i},
\notag\\
\Xi_{e\gamma}^{\max}
&\coloneqq w_N^\dagger(D_i^{(0)})^{-1}w_N.
\label{eq:projection-bound}
\end{align}
At $M_i=1\,\mathrm{TeV}$,
\begin{equation}
\Xi_{e\gamma}^{\max}=\XiGeneral,
\qquad
\rho_\star=\RhoStar.
\label{eq:projection-maximum-numeric}
\end{equation}
Equality requires every nonzero $z_{\alpha i}$ to be proportional to
$(D_i^{(0)})^{-1}w_N$.  The bound maximizes over electroweak directions
allowed by the physical width and is not the value of the factorized
benchmark.

\begin{figure}
\includegraphics[width=\columnwidth]{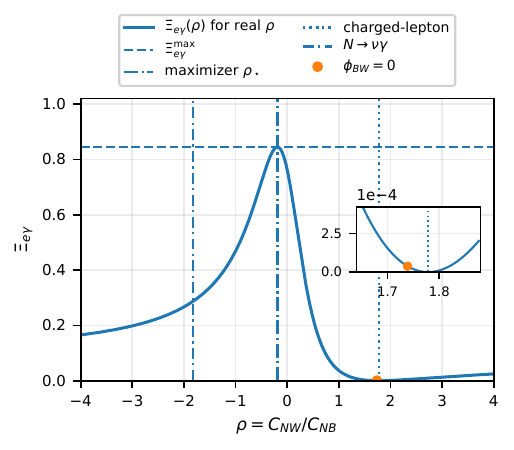}
\caption{Width-normalized charged-lepton photon projection for real
$\rho=C_{NW}/C_{NB}$ at $M=1\,\mathrm{TeV}$.  The horizontal dashed line
is the maximum over complex electroweak directions.  The vertical dash-dotted
line marks the maximizer $\rho_\star$, the dotted line marks the charged-lepton
blind direction, and the long-dashed line marks the $N\to\nu\gamma$ blind
direction.  The circular marker shows the benchmark $\phi_{BW}=0$,
for which $\rho=+\RhoBWBenchmark$; the inset resolves its distance from
$\rho=c_N/s_N$.}
\label{fig:projection}
\end{figure}

In the common-ratio slice $z\propto(1,\rho)^{\intercal}$, the width-normalized
charged-lepton photon projection is
\begin{align}
\Xi_{e\gamma}(\rho;M)
&\coloneqq\frac{|c_N-s_N\rho|^2}{D_0(\rho;M)},
\label{eq:Xi-rho}\\
D_0(\rho;M)
&\coloneqq|c_0+s_0\rho|^2
+R_Z^{\rm vac}|-s_0+c_0\rho|^2
\notag\\[-2pt]
&\hspace{25mm}+2R_W^{\rm vac}|\rho|^2.
\end{align}
For the benchmark phase $\phi_{BW}=0$, so that $\rho=+\RhoBWBenchmark$,
$\Xi_{e\gamma}=\XiReportedReal$.  If only the modulus
$|\rho|=\RhoBWBenchmark$ is
retained and the phase is relaxed, maximizing
over the relative phase gives $\Xi_{e\gamma}=\XiReportedPhase$ at
$\phi_{BW}=\pi$.  The relaxed-phase maximum is a sensitivity slice, not a
benchmark prediction.  Figure~\ref{fig:projection} shows $\Xi_{e\gamma}(\rho)$ for real
$\rho$ together with the maximizer and the two blind directions.

\section{LEFT evolution and charged-lepton observables}
\label{sec:observables}

\subsection{QED evolution}

The operator in Eq.~\eqref{eq:LEFT-operator} contains no explicit factor of
$e$.  In an interval with a fixed set of charged Dirac fermions, define
\begin{equation}
S_f(\kappa)\coloneqq\sum_{f\,{\rm active}}N_c^f q_f^2.
\end{equation}
The one-loop LEFT equations are \cite{JenkinsManoharStofferADM}
\begin{align}
\kappa\frac{\dd\alpha}{\dd\kappa}
&=\frac{2S_f}{3\pi}\alpha^2,
\label{eq:qed-alpha}\\
\kappa\frac{\dd(C_{e\gamma})_{\alpha\beta}}{\dd\kappa}
&=\frac{\alpha}{4\pi}
\left(10+\frac43S_f\right)(C_{e\gamma})_{\alpha\beta}.
\label{eq:qed-dipole}
\end{align}
For a fixed active set,
\begin{equation}
\frac{C_{e\gamma}(\kappa)}{C_{e\gamma}(\kappa_0)}
=\left[\frac{\alpha(\kappa)}{\alpha(\kappa_0)}\right]^{(15+2S_f)/(4S_f)}.
\label{eq:qed-solution}
\end{equation}
The stepwise prescription in Appendix~\ref{app:inputs} gives
\begin{align}
C_{e\gamma}(m_\mu)&=\QEDMuon\,C_{e\gamma}(\kEW),
\notag\\
C_{e\gamma}(m_e)&=\QEDElectron\,C_{e\gamma}(\kEW).
\label{eq:qed-factors}
\end{align}
Below the hadronic scale, current-quark masses are used only as a
leading-logarithmic interpolation.  Removing $u,d,s$ collectively at
$1\,\mathrm{GeV}$ instead gives $\QEDMuonAlt$ and $\QEDElectronAlt$, shifts of
$\QEDMuonShift\%$ and $\QEDElectronShift\%$, respectively.  The auxiliary
running values of $\alpha$ below confinement are not interpreted as physical
measurements of the QED coupling.

The factors in Eq.~\eqref{eq:qed-factors} are applied as a universal RG
improvement of the linear coefficient.

\subsection{Observable normalizations and present limits}

With the Lagrangian convention in Eq.~\eqref{eq:LEFT-operator}, the
dipole-observable relations are \cite{Aebischer2021,ZhangZhouRadiative2021}
\begin{align}
\mathrm{BR}(\mu\to e\gamma)
={}&\frac{48\pi^2}{G_F^2m_\mu^2}
\Bigl[|(C_{e\gamma})_{e\mu}(m_\mu)|^2
\notag\\[-2pt]
&\hspace{23mm}+|(C_{e\gamma})_{\mu e}(m_\mu)|^2\Bigr],
\label{eq:BR-muegamma}\\
d_e={}&2\,\mathrm{Im}(C_{e\gamma})_{ee}(m_e),
\label{eq:electron-edm}\\
\Delta a_\mu={}&-\frac{4m_\mu}{e(m_\mu)}
\mathrm{Re}(C_{e\gamma})_{\mu\mu}(m_\mu).
\label{eq:muon-gminus2}
\end{align}
A coefficient in $\mathrm{GeV}^{-1}$ is converted to $e\,\mathrm{cm}$ in
Eq.~\eqref{eq:electron-edm} by multiplying by
$1.973269804\times10^{-14}\,\mathrm{cm}/e(0)$, where
$e(\kappa)\coloneqq\sqrt{4\pi\alpha(\kappa)}$ and $\alpha(\kappa)$ is the
running fine-structure constant.

The current limits are
\begin{align}
\mathrm{BR}(\mu^+\to e^+\gamma)&<1.5\times10^{-13},
&&90\%\ \mathrm{C.L.},
\label{eq:MEG-limit}\\
|d_e|&<4.1\times10^{-30}\,e\,\mathrm{cm},
&&90\%\ \mathrm{C.L.},
\label{eq:EDM-limit}
\end{align}
from MEG II \cite{MEGII2025} and the HfF$^+$ molecular measurement
\cite{Roussy2023}.  The EDM interpretation assumes that the electron dipole
dominates the measured CP-odd molecular response.

The final experimental world average of the muon anomaly has uncertainty
\cite{MuonGminus2FinalReport}
\begin{equation}
\sigma_{a_\mu}^{\rm exp}=145\times10^{-12}.
\label{eq:amu-uncertainty}
\end{equation}
The 2025 theory update finds no statistically significant discrepancy when
its lattice-QCD HVP average is used, while tensions among dispersive HVP data
sets remain \cite{MuonTheory2025}.  We therefore quote $|\Delta a_\mu|$ and
its ratio to Eq.~\eqref{eq:amu-uncertainty}, rather than selecting a unique
new-physics interval.

\section{Results}
\label{sec:results}

\subsection{Conditional envelope}

For one heavy state, Eqs.~\eqref{eq:Y-norm-bound} and
\eqref{eq:projection-bound} imply
\begin{align}
|(Y_\nu)_{\beta i}|
&\leqslant\frac{2\sqrt{\epsilon M_i\mEM_i}}{v_0},
\label{eq:Y-entry-bound}\\
|\mu^{e\gamma}_{\alpha i}|
&\leqslant\sqrt{\Xi_{e\gamma}^{\max}(M_i)}
\frac{\sqrt{\mEM_i}}{v_0\sqrt{M_i}}.
\label{eq:mu-entry-bound}
\end{align}
Combining these inequalities in Eq.~\eqref{eq:P-definition} and aligning the
heavy-state phases gives
\begin{equation}
|P_{\alpha\beta}|
\leqslant P_{\rm env}
\coloneqq\frac{2\sqrt\epsilon}{v_0^2}
\sum_i L_i\sqrt{\Xi_{e\gamma}^{\max}(M_i)}\,\mEM_i.
\label{eq:P-envelope}
\end{equation}
The explicit powers of $M_i$ cancel; residual dependence remains through
$L_i$, $R_Z$, and $R_W$.

Equation~\eqref{eq:P-envelope} maximizes one matrix element.  The exact global
maximum of $\mu\to e\gamma$ concentrates the dipole norm in the electron row
and the Yukawa norm in the muon row, so only
$(C_{e\gamma})_{e\mu}$ is nonzero at the maximum.  The opposite chirality
cannot increase the result because it carries $y_e$ rather than $y_\mu$.
The EDM and $g-2$ maxima require different flavor and phase assignments; the
three envelopes below are separately valid but are not generally saturated
by one common texture.

For the companion factorized dipoles,
\begin{equation}
|\mu^{e\gamma}_{\alpha i}|^2
=f_\alpha\,\Xi_{e\gamma}(\rho_{BW};M_i)
\frac{\mEM_i}{v_0^2M_i},
\label{eq:factorized-projection}
\end{equation}
with $\Xi_{e\gamma}(\rho;M)$ defined in Eq.~\eqref{eq:Xi-rho}.
This retains the dipole flavor fractions and the electroweak ratio while still
optimizing the unspecified $Y_\nu$ under Eq.~\eqref{eq:Y-norm-bound}.

\subsection{Benchmark-scale numbers}

We use the conditional reference point
\begin{equation}
\begin{gathered}
\epsilon=10^{-2},\qquad M_{1,2}=1\,\mathrm{TeV},\\[-2pt]
\mEM_1=\mEM_2=\MtildeBenchmark\,\mathrm{eV},
\end{gathered}
\label{eq:numerical-reference}
\end{equation}
with constructive alignment of the two heavy-state contributions.  The value
of $\epsilon$ is not an input of the companion benchmark.  For each Yukawa
column,
\begin{equation}
\|Y_i\|_2\leqslant\YNormBound.
\label{eq:Y-norm-numeric}
\end{equation}

\begin{table*}[t]
\caption{QED-evolved observables obtained from the linear one-loop
leading-logarithmic coefficient at the reference point in
Eq.~\eqref{eq:numerical-reference}.  Each column is optimized separately.}
\label{tab:results}
\begin{ruledtabular}
\begin{tabular}{lccc}
Assumptions
& $\mathrm{BR}(\mu\to e\gamma)$
& $|d_e|\;(e\,\mathrm{cm})$
& $|\Delta a_\mu|$ \\
\hline
Width-only optimized envelope
& $\GeneralBR$ & $\GeneralEDM$ & $\GeneralAMu$ \\
Factorized flavor structure, $|\rho_{BW}|=\RhoBWBenchmark$, phase optimized
& $\FixedBR$ & $\FixedEDM$ & $\FixedAMu$ \\
Factorized flavor structure, $\phi_{BW}=0$
($\rho_{BW}=+\RhoBWBenchmark$)
& $\RealBR$ & $\RealEDM$ & $\RealAMu$
\end{tabular}
\end{ruledtabular}
\end{table*}

The first row of Table~\ref{tab:results} optimizes arbitrary electroweak and flavor directions.  The
second retains the factorized fractions and fixed modulus but relaxes the
benchmark phase and maximizes over it.  The third uses the benchmark phase
$\phi_{BW}=0$ and is therefore
sensitive to the nearby blind direction.  Because the companion benchmark does
not specify $Y_\nu$, none of the rows is fixed by that benchmark alone.

For the width-only envelope, the sensitivity ratios are
\begin{align}
\frac{\mathrm{BR}(\mu\to e\gamma)}{1.5\times10^{-13}}
&\leqslant\GeneralBRRatio,
\notag\\
\frac{|d_e|}{4.1\times10^{-30}\,e\,\mathrm{cm}}
&\leqslant\GeneralEDMRatio,
\notag\\
\frac{|\Delta a_\mu|}{\sigma_{a_\mu}^{\rm exp}}
&\leqslant\GeneralAMuRatio.
\label{eq:benchmark-ratios}
\end{align}
The electron-EDM envelope gives the largest ratio, approximately
$1.4\times10^{-7}$.

The $\phi_{BW}=0$ row is the benchmark evaluation of the linear logarithm, not a
fixed-order prediction.  Its suppression is caused by
$|c_W-s_W\rho_{BW}|$.  The displayed inputs are rounded.
Changing the common weak-angle scale is a higher-order effect whose relative
impact is enhanced near the zero, while finite threshold terms need not share
the same cancellation.

\subsection{Dependence on \texorpdfstring{$\mEM$, $\epsilon$, and $\kmatch$}{the width, epsilon, and the matching scale}}

\begin{figure}[tb]
\includegraphics[width=\columnwidth]{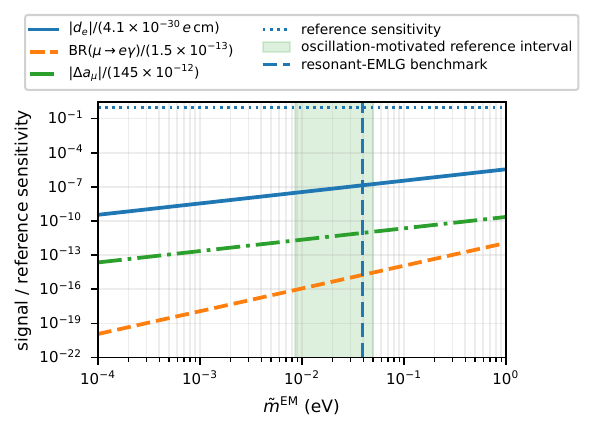}
\caption{Two-state, coherently aligned width-only linear envelopes for
$\epsilon=10^{-2}$, $M_{1,2}=1\,\mathrm{TeV}$, and
$\kmatch=3\,\mathrm{TeV}$.  The shaded band is the oscillation-motivated
reference interval used in the companion transport scan, and the vertical
dashed line marks $\mEM_1=\MtildeBenchmark\,\mathrm{eV}$.  The curves are
normalized to the present $d_e$ and $\mu\to e\gamma$ limits and to the final
experimental uncertainty of $a_\mu$.}
\label{fig:width-envelope}
\end{figure}

At fixed heavy mass and electroweak direction,
\begin{align}
|C_{e\gamma}^{\rm lin,LL}|&\propto
\sqrt\epsilon\,\mEM\ln\frac{\kmatch}{M},
\notag\\
\mathrm{BR}(\mu\to e\gamma)&\propto
\epsilon(\mEM)^2\ln^2\frac{\kmatch}{M},
\notag\\
|d_e|,\ |\Delta a_\mu|&\propto
\sqrt\epsilon\,\mEM\ln\frac{\kmatch}{M}.
\label{eq:scalings}
\end{align}
Figure~\ref{fig:width-envelope} shows the width-only linear envelope across
the $10^{-4}$--$1\,\mathrm{eV}$ range displayed in the companion scan.  The
shaded $8.6\times10^{-3}$--$5.0\times10^{-2}\,\mathrm{eV}$ interval is an
oscillation-motivated reference range, not a direct bound on $\mEM$.  The pole
and observed-yield crossing locations are listed separately in
Table~\ref{tab:poles}.

At $\mEM_1=\mEM_2=1\,\mathrm{eV}$, the same linear envelope gives
\begin{align}
\mathrm{BR}(\mu\to e\gamma)&\leqslant\EndpointBR,
\notag\\[-2pt]
\frac{\mathrm{BR}}{\mathrm{BR}_{\rm lim}}&\leqslant\EndpointBRRatio,
\notag\\
|d_e|&\leqslant\EndpointEDM\,e\,\mathrm{cm},
\notag\\[-2pt]
\frac{|d_e|}{|d_e|_{\rm lim}}&\leqslant\EndpointEDMRatio,
\notag\\
|\Delta a_\mu|&\leqslant\EndpointAMu,
\notag\\[-2pt]
\frac{|\Delta a_\mu|}{\sigma_{a_\mu}^{\rm exp}}&\leqslant\EndpointAMuRatio.
\label{eq:endpoint-results}
\end{align}

\begin{figure}[tb]
\includegraphics[width=\columnwidth]{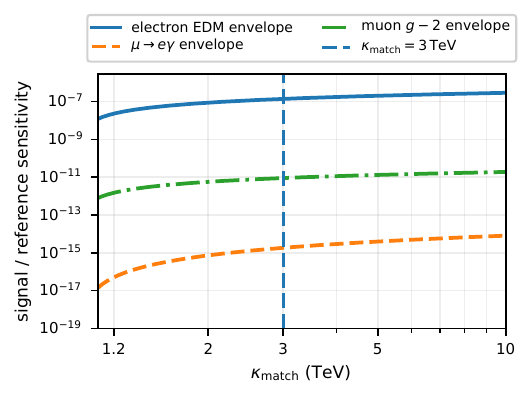}
\caption{Matching-scale dependence of the two-state width-only linear
envelopes at $\mEM_i=\MtildeBenchmark\,\mathrm{eV}$ and $\epsilon=10^{-2}$.
The curves arise from $L_i=\ln(\kmatch/M_i)$; the vertical dashed line marks
the companion value $\kmatch=3\,\mathrm{TeV}$.  Finite threshold matching and
RGE insertions inside the one-loop source are not included.}
\label{fig:scale-envelope}
\end{figure}

Figure~\ref{fig:scale-envelope} varies $\kmatch$ while holding the threshold
inputs and physical width fixed.  The curves show only the leading-logarithmic
scale dependence and are not a complete theory-uncertainty estimate.

\section{Discussion and conclusions}
\label{sec:discussion}

The resonant EMLG benchmark fixes the neutrino-dipole sector but not the
renormalizable matrix $Y_\nu$.  The charged-lepton dipole term calculated here
is linear in $C_{NX}$ and $Y_\nu$ and is evaluated after imposing the width
bound in Eq.~\eqref{eq:width-inequality} and the UV boundary condition in
Eq.~\eqref{eq:direct-boundary}.

At $\epsilon=10^{-2}$, the QED-evolved width-only linear envelopes are below
the reference experimental sensitivities across the width range in
Fig.~\ref{fig:width-envelope}.  At the physical-width benchmark, the largest
ratio is $|d_e|/|d_e|_{\rm lim}\leqslant\GeneralEDMRatio$.  The factorized
flavor structure and
the fixed electroweak modulus reduce the conditional envelopes.  The much smaller value at the benchmark $\phi_{BW}=0$ is not a precision prediction because it lies near the charged-lepton projection zero.

The leading-logarithmic result is not the complete fixed-order one-loop linear
amplitude.  Finite matching at the heavy-neutrino threshold is omitted, and the
LEFT QED evolution is included separately as an RG improvement.  This
limitation is most relevant at the benchmark $\phi_{BW}=0$, where the retained
projection is small.

The thermal pole prescription and vacuum local matching enter different
parts of the analysis.  The two pole prescriptions select different locations
for the positive maximum and observed-yield crossing in the transport
calculation.  The local coefficient is analytic in the quasi-degenerate masses
and changes by only $\order(\Delta M/M_1)$ between the two selected crossings.

The width bound also constrains the type-I mass term:
\begin{equation}
\|m_\nu^{(I)}\|_2
\leqslant\frac{v_0^2}{2}\sum_i\frac{\|Y_i\|_2^2}{M_i}
\leqslant2\epsilon\sum_i\mEM_i.
\label{eq:type-I-bound}
\end{equation}
At the two-state reference point,
$\|m_\nu^{(I)}\|_2\leqslant\TypeIBound\,\mathrm{eV}$.  This is below the atmospheric
mass scale near $0.05\,\mathrm{eV}$ in current oscillation fits
\cite{NuFit60,NuFIT61}.  If the light-neutrino masses are Majorana and are to be explained within this
setup, the type-I term alone is insufficient and another mass-generating
contribution is required.  Oscillation data alone do not establish that the
light neutrinos are Majorana.  Dipole-induced radiative masses impose further
UV-dependent conditions \cite{BellKayserLaw2008,BellMajorana2006,BellNaturalness2005}
and are not calculated here.  KATRIN gives the direct bound
$m_\beta<0.45\,\mathrm{eV}$ at $90\%$ confidence \cite{KATRIN2025}.

The result of this paper is therefore a conditional bound on the specified
Yukawa-assisted linear portal.  Within its stated boundary conditions, present
experimental sensitivities remain many orders of magnitude above the
calculated envelopes.

\appendix

\section{Two-body width normalization}
\label{app:widths}

Consider the broken-phase interaction
\begin{equation}
-\mathcal L\supset
\mu^V\bar f\sigma^{\mu\nu}P_RN V_{\mu\nu}+\mathrm{h.c.},
\label{eq:app-interaction}
\end{equation}
with $m_f=0$, $P^2=M^2$, $q^2=m_V^2$, and $r=m_V^2/M^2$.  For one charge
channel,
\begin{equation}
\mathcal M=2\mu^V\bar u(p)\sigma^{\mu\nu}q_\mu
\varepsilon_\nu^{\ast}(q)P_Ru(P).
\end{equation}
Averaging over the initial spin and summing over final spins and
polarizations gives
\begin{equation}
\frac12\sum|\mathcal M|^2
=4|\mu^V|^2M^4(1-r)\left(1+\frac r2\right).
\end{equation}
The $q_\nu q_\rho/m_V^2$ term in the polarization completeness relation drops
out because $q_\mu q_\nu\sigma^{\mu\nu}=0$; this does not remove the physical
longitudinal polarization contribution contained in the remaining tensor
contraction.  The width of one charge channel is
\begin{equation}
\Gamma
=\frac{M^3}{4\pi}|\mu^V|^2(1-r)^2
\left(1+\frac r2\right).
\end{equation}
For a Majorana field, Eq.~\eqref{eq:app-interaction} and its Hermitian
conjugate produce equal charge-conjugate channels.  Their sum yields
Eq.~\eqref{eq:majorana-width}.

For the weak operator in Eq.~\eqref{eq:ONW},
$\tau^1\tilde H$ and $\tau^2\tilde H$ combine into
$W^1+\mathrm{i}W^2=\sqrt2W^-$.  Therefore
$\mu^W=v_0C_{NW}=\sqrt2\mu^3$, and the physical-width quadratic form contains
$2R_W^{\rm vac}|\mu^3|^2$.

For the Yukawa interaction, one charge channel contributes
$M(Y_\nu^\dagger Y_\nu)_{ii}/(16\pi)$.  Summing the two charge-conjugate
channels gives Eq.~\eqref{eq:yukawa-width} and the factor four in
Eq.~\eqref{eq:Y-norm-bound}.

\section{Projection and flavor-envelope saturation}
\label{app:envelope}

In the $D_i^{(0)}$-weighted inner product,
\begin{equation}
|w_N^{\intercal}z|^2
\leqslant
\bigl[w_N^\dagger(D_i^{(0)})^{-1}w_N\bigr]
\bigl[z^\dagger D_i^{(0)}z\bigr].
\end{equation}
This proves Eq.~\eqref{eq:projection-bound}.  Equality requires
$z\propto(D_i^{(0)})^{-1}w_N$.  In the common-ratio slice
$z\propto(1,\rho)^{\intercal}$, the equality vector gives the real ratio
$\rho=\rho_\star$ in Eq.~\eqref{eq:projection-maximum-numeric}.

For each heavy state,
\begin{equation}
|(Y_\nu)_{\beta i}^{\ast}\mu^{e\gamma}_{\alpha i}|
\leqslant
\frac{2\sqrt\epsilon}{v_0^2}
\sqrt{\Xi_{e\gamma}^{\max}(M_i)}\,\mEM_i.
\end{equation}
The triangle inequality over the heavy states gives
Eq.~\eqref{eq:P-envelope}.  Saturation requires simultaneous saturation of
the Yukawa-width inequality and the electroweak projection bound, allocation
of the allowed flavor norms to the selected matrix element, and constructive
heavy-state phases.

For the flavor-changing rate, let the dipole and Yukawa flavor vectors of one
state be $a_\alpha$ and $b_\beta$, with
$\sum_\alpha|a_\alpha|^2\leqslant A^2$ and
$\sum_\beta|b_\beta|^2\leqslant B^2$.  Then
\begin{equation}
|C_{e\mu}|^2+|C_{\mu e}|^2
\propto y_\mu^2|a_e|^2|b_\mu|^2
+y_e^2|a_\mu|^2|b_e|^2
\leqslant y_\mu^2A^2B^2.
\end{equation}
The maximum is reached by $a_e=A$, $b_\mu=B$, with the other components zero.
Separately maximizing both chiralities and adding them in quadrature would
therefore not describe one saturating texture.  For the factorized dipole
flavor structure, the same ordering follows from
$f_ey_\mu^2\gg f_\mu y_e^2$.

\section{Numerical inputs, coupled running, and QED thresholds}
\label{app:inputs}

Table~\ref{tab:inputs} lists the numerical inputs.  The SM couplings at
$\kEW=150\,\mathrm{GeV}$ are the common inputs of the companion benchmark and
are evolved at one loop for the RGE and strict-LL projection.  To reproduce its
zero-temperature width normalization, the phase-space masses are
$m_W=g(\kEW)v_0/2$ and
$m_Z=\sqrt{g^2(\kEW)+g'^2(\kEW)}\,v_0/2$.  They are benchmark tree-level
quantities rather than precision pole-mass inputs.  Charged-lepton and quark
masses enter Yukawa couplings, observable normalizations, and the stepwise QED
interpolation.  Standard fermion masses and $G_F$ follow the Particle Data
Group compilation \cite{PDG2024}.

\begin{table}[tb]
\caption{Principal numerical inputs.}
\label{tab:inputs}
\begin{ruledtabular}
\begin{tabular}{ll}
Quantity & Value \\
\hline
$\kmatch$ & $3\,\mathrm{TeV}$ \\
$\kN=M_{1,2}$ & $1\,\mathrm{TeV}$ \\
$\kEW$ & $150\,\mathrm{GeV}$ \\
$\epsilon$ & $10^{-2}$ \\
$v_0$ & $246\,\mathrm{GeV}$ \\
$G_F$ & $1.1663787\times10^{-5}\,\mathrm{GeV}^{-2}$ \\
$m_e$ & $0.510998950\,\mathrm{MeV}$ \\
$m_\mu$ & $105.6583755\,\mathrm{MeV}$ \\
$m_W$ (width normalization) & $\MWWidth\,\mathrm{GeV}$ \\
$m_Z$ (width normalization) & $\MZWidth\,\mathrm{GeV}$ \\
$g'(150\,\mathrm{GeV})$ & $0.357$ \\
$g(150\,\mathrm{GeV})$ & $0.652$ \\
$g_s(150\,\mathrm{GeV})$ & $1.17$ \\
$y_t(150\,\mathrm{GeV})$ & $0.93$ \\
$\alpha(150\,\mathrm{GeV})$ & $1/127.95$
\end{tabular}
\end{ruledtabular}
\end{table}

For the RGE translation in Eq.~\eqref{eq:RGE-transfer}, the auxiliary
one-loop SM equations are
\begin{align}
16\pi^2\frac{\dd g'}{\dd\ln\kappa}
&=\frac{41}{6}g'^3,\quad
16\pi^2\frac{\dd g}{\dd\ln\kappa}
=-\frac{19}{6}g^3,
\notag\\
16\pi^2\frac{\dd g_s}{\dd\ln\kappa}
&=-7g_s^3,
\notag\\
16\pi^2\frac{\dd y_t}{\dd\ln\kappa}
&=y_t\left(\frac92y_t^2-\frac{17}{12}g'^2-\frac94g^2-8g_s^2\right).
\end{align}
Applying the real transfer matrix in Eq.~\eqref{eq:RGE-transfer} to the
unrounded hard-scale coefficients gives the phase-dependent matching-scale
magnitudes
\begin{align}
\phi_{BW}=0:&\quad(\RGEPositiveNB,\RGEPositiveNW)\,\mathrm{GeV}^{-2},
\notag\\
\phi_{BW}=\frac{\pi}{2}:&\quad(\RGEHalfPiNB,\RGEHalfPiNW)\,\mathrm{GeV}^{-2},
\notag\\
\phi_{BW}=\pi:&\quad(\RGEPiNB,\RGEPiNW)\,\mathrm{GeV}^{-2}.
\end{align}
These examples quantify the sensitivity of the endpoint magnitudes to the
relative phase away from the benchmark value $\phi_{BW}=0$.

For Eq.~\eqref{eq:qed-factors}, charged fermions are decoupled stepwise at
$m_b=4.18\,\mathrm{GeV}$, $m_\tau=1.77686\,\mathrm{GeV}$,
$m_c=1.27\,\mathrm{GeV}$, $m_\mu=105.6583755\,\mathrm{MeV}$,
$m_s=93\,\mathrm{MeV}$, $m_d=4.67\,\mathrm{MeV}$,
$m_u=2.16\,\mathrm{MeV}$, and $m_e=0.510998950\,\mathrm{MeV}$.
The $u,d,s$ thresholds below confinement serve only as an interpolation.
Decoupling all three light quarks at $1\,\mathrm{GeV}$ gives the alternative
factors quoted in Sec.~\ref{sec:observables}.

\end{document}